\documentclass[aip,pop,onecolumn,preprintnumbers,amssymb]{revtex4-1}

\usepackage{amsmath}
\usepackage{graphicx}% Include figure files
\usepackage{dcolumn}% Align table columns on decimal point
\usepackage{bm}% bold math

\begin{document}

\draft

\title{A Parametric Investigation on the Cyclotron Maser Instability Driven by Ring-beam Electrons with Intrinsic Alfv\'en Waves}

\author{Zi-Jin Tong}
\affiliation{CAS Key Laboratory of Geospace Environment,
Schoool of Earth and Space Sciences, University of
Science and Technology of China, Hefei, Anhui 230026, China}
\author{Chuan-Bing Wang}
\email{cbwang@ustc.edu.cn, corresponding author.}
\affiliation{CAS Key Laboratory of Geospace Environment,
Schoool of Earth and Space Sciences, University of
Science and Technology of China, Hefei, Anhui 230026, China}
\affiliation{Collaborative Innovation Center of Astronautical Science and Technology, China}
\author{Pei-Jin Zhang}
\affiliation{CAS Key Laboratory of Geospace Environment,
Schoool of Earth and Space Sciences, University of
Science and Technology of China, Hefei, Anhui 230026, China}
\author{Jin Liu}
\affiliation{CAS Key Laboratory of Geospace Environment,
Schoool of Earth and Space Sciences, University of
Science and Technology of China, Hefei, Anhui 230026, China}
\affiliation{School of Resource Environment and Earth Science, Yunnan University, Kunming, Yunan 650091, China}

\date{\today}

\begin{abstract}
The electron-cyclotron maser is a process that generates intense and coherent radio emission in plasma. In this paper, we present a comprehensive parametric investigation on the electron-cyclotron-maser instability driven by non-thermal ring-beam electrons with intrinsic Alfv\'en waves which pervade the solar atmosphere and interplanetary space. It is found that both forward propagating and backward propagating waves can be excited in the fast ordinary (O) and extraordinary (X) electromagnetic modes. The growth rates of X1 mode are almost always weakened by Alfv\'en waves. The average pitch-angle $\phi_0$ of electrons is a key parameter for the effect of Alfv\'en waves on the growth rate of modes O1, O2 and X2. For a beam-dominated electron distribution ($\phi_0 \lesssim 30^\circ$ ), the growth rates of the maser instability for O1, O2 and X2 modes are enhanced with the increase of Alfv\'en wave energy density. In other conditions, the growth rates of O1, O2 and X2 modes weakened with increasing Alfv\'en wave intensity, except that the growth of O1 mode may also be enhanced by Alfv\'en waves for a ring distribution. The results may be important for us in analyzing the mechanism of radio bursts with various fine structures observed in space and astrophysical plasmas.
\end{abstract}

\maketitle

\section{Introduction}

Electron-cyclotron-maser (ECM) instability is an appealing mechanism for direct amplification of radio radiation by non-thermal (energetic) electrons, which has been widely applied for explanations of high-power radio emissions from magnetized planets as well as other astronomical objects\cite{Treumann2006}, and for microwave generation in the laboratory \cite{Chu2004}. Twiss \cite{Twiss1958} and Schneider \cite{Schneider1959} first pointed out independently that this induced emission mechanism can amplify electromagnetic wave at radio frequency which is close to the electron-cyclotron frequency and its harmonics in a magnetic field. Hirshfield and Wachtel \cite{Hirshfield1964} demonstrated the ECM emission by microwave generation in a gyrotron with relativistic electrons in the laboratory. Within the space and astrophysical context, the ECM instability did not receive intensive attentions until the late 1970s when the Earth's auroral kilometric radiation (AKR) was successfully explained by Wu and Lee \cite{WuLee1979} in terms of the cyclotron-maser mechanism. They realized that including the weakly relativistic effect could dramatically increase the efficiency of the amplification of electromagnetic wave. Since then, the ECM instability has been extensively investigated in the 1980s, and has gained increasing attention in the last three decades with its applications to various radio emissions beyond the Earth, for example, the solar radio bursts \cite{Melrose1982,Vlahos1987, Wu2002,Wu2005,Zhao2013,Wang2015,Melrose2016}, the radio emission from magnetized planets in our solar system \cite{Zarka1998, Gurnett2004, Menietti2010} and the time-varying emission from magnetized blazar jets \cite{Begelman2005}. One can refer to the review literatures \cite{Wu1985, Treumann2006, Bingham2013} for more details about the fundamentals of this instability and its applications in astrophysics.

There are two necessary conditions for the excitation of ECM instability. The first condition is that the local electron plasma frequency is less than the electron-cyclotron frequency for significant excitation of the fast ordinary and extraordinary electromagnetic waves. This indicates that the background plasma density may be depleted in the radio source region as demonstrated by \textit{in situ} observation in the source region of AKR \cite{Benson1980}. There are only few works published on the physical process for this density depletion. It is considered that the plasma is diluted by a magnetic-field-aligned electric potential drop \cite{Lyons1980, Vedin2005, Treumann2006} for the Earth's AKR. For a plasma with low plasma-beta value such as in the solar corona, the density may be depleted due to magnetic compression or Alfv\'en waves excited in a magnetic flux tube by beams of energetic particles through pressure balance across the boundary of the flux tube \cite{Wu2006, Wu2014}.

The second condition is that the electrons possess a velocity distribution with a perpendicular population inversion which provides the energy to amplify the electromagnetic wave via wave-particle resonant interaction. A number of velocity-distribution models have been proposed and studied for cyclotron maser that occurs in different physical environments, such as the loss-cone distribution (or a ring distribution, the most famous one)\cite{Dory1965, WuLee1979, Lee1980}, the ring-shell distribution \cite{Bekefi1961, Wu2008}, the horseshoe distribution (or partial-shell distribution) \cite{Winglee1986, Vorgul2011} and the hollow ring-beam distribution \cite{Chu1978, Yoon2007}. Basically, the electron velocity distribution for maser instability can be divided into two components, namely, a ring and a beam. The ring is the velocity component perpendicular to the ambient magnetic field, which is the energy source for the cyclotron-maser instability. The beam is the velocity component parallel to the ambient magnetic field line. The relative amplitude and distribution of these two components determine the pattern and position of the area where the distribution function has positive gradient $\partial F_e/\partial v_\perp$ in phase space, which further gives the wave modes that are unstable and correspondingly their growth rates, frequencies and wave propagation angles, because the resonance ellipse of an unstable wave mode should be located inside the region where $\partial F_e/\partial v_\perp>0$. Here $F_e$  and $v_\perp$ are the electron velocity distribution function and the perpendicular velocity with respect to the ambient magnetic field.

It is well-known that large-amplitude intrinsic Alfv\'en waves are observed in the solar atmosphere and interplanetary space \cite{Belcher1971, Pontieu2007, McIntosh2011, Li2016}. Much work has been published discussing the roles of Alfv\'en waves in the heating and acceleration of solar corona and solar wind through resonant or non-resonant wave-particle interaction \cite{Marsch1982,Hollweg2002,Isenberg2011,Wang2014}. Recently, it is found that the pre-existing Alfv\'en waves enhance the ECM instability \cite{Wu2012} driven by non-thermal electrons with a beam-feature distribution. The waves qualitatively affect the velocity distribution of energetic electrons via pitch-angle scattering. A beam distribution function can deform quickly to become a crescent-shaped distribution so that the cyclotron-maser instability can be excited. At the same time the intrinsic Alfv\'en waves may modify the classical cyclotron resonance processes.

In this paper, we discuss the influences of intrinsic Alfv\'en waves on the growth rate of cyclotron-maser instability driven by non-thermal ring-beam electrons. Comparing with previously theoretical work, in most of which the growth rates were calculated for a few special parameter values, a comprehensive parametric investigation on the dependence of growth rates on several parameters is done in the present paper so that we can gain a more complete view on the maser excitation at different physical conditions. The results can provide important clues for us to explore the various fine features of radio emission observed in nature.

The organization of the paper is as follows. In Section 2, the model of the scattered ring-beam electron distribution function is introduced and the general formulae for the growth rate is described. In Section 3, some sample growth rate calculations are presented to show the general characteristics of the maser instability. In Section 4, we display the results from the comprehensive parametric investigation on the maximum growth rate. Finally, we conclude in Section 5.

\section{Model of the distribution function and formulae of the growth rate}

\subsection{Distribution of ring-beam electrons scattered by Alfv\'en waves}

The ring-beam distribution is not an infrequent feature for non-thermal (energetic) electrons accelerated in space and astrophysical plasmas. For instance, based on the physics of collisionless shock waves, the reflected electrons from quasi-perpendicular shock layer is characterized by ring-beam or loss-cone-beam distribution \cite{Wu1984, Farrell2001}, which can be the energy source for radio emission from collisionless shock \cite{Bingham2003, Yoon2007, Zhao2014}. Solar type III radio bursts are produced by fast electron beams along open magnetic field line \cite{Wild1954, Robinson1998, Reid2014, Wang2015}, which may also have a ring component if they are initially injected to the open field with an angle with respect to ambient field line from the magnetic reconnection site.

When there are pre-existing Alfv\'en waves, these waves can pitch-angle scatter the electrons. During the pitch-angle scattering, the electron energy is conserved in a frame moving with the Alfv\'en waves. If the beam component is the dominant velocity for the non-thermal electron without Alfv\'en waves, the pitch-angle scattering will mainly accelerate the electrons in the transverse direction and will increase the ring component. On the other hand, if the ring component is the dominant component without Alfv\'en waves, the pitch-angle scattered electrons will mainly spread their pitch-angle distribution. As a result, a partial-shell (or crescent-shaped) distribution emerges, and a spherical shell (or ring-shell) distribution may develop, depending on the intensity of the Alfv\'en waves. Introducing the momentum per unit mass of rest electron $\bold{u}=\bold{p}/m_{e0}$, we model the scattered ring-beam electron distribution simply as follows.
%(1)
\begin{equation}
F_e(u,\mu)=A\exp{\left[-\frac{(u-u_0)^2}{\alpha^2}-\frac{(\mu-\mu_0)^2}{\beta^2}\right]},
\end{equation}
where $\mu=\cos\phi=u_{\parallel}/u$, and $\phi$ is the pitch angle, $u_{\parallel}$ is the momentum parallel to the ambient magnetic field. The quantities, $u_0$ and $\mu_0$ represent the average electron momentum and the cosine of the average pitch-angle $\phi_0$, $\alpha$ and $\beta$  designate the momentum dispersion and the spread width in pitch angle, respectively. The value of $\beta$ is proportional to the normalized energy density of Alfv\'en waves. On the basis of a preceding analytical study \cite{Yoon2009} we assume that $\beta^2\approx2B_w^2/B_0^2$, where $B_w$ and $B_0$ are the strength of the Alfv\'en wave magnetic field and the background magnetic field, respectively. The normalization constant $A$ is given by
%(2)
\begin{equation}
\frac{1}{A}=\frac{\pi^{3/2}}{\sqrt{2}}\beta\alpha^3e^{-u_0^2/2\alpha^2}\textrm{D}_{-3}\left(-\frac{\sqrt{2}u_0}{\alpha}\right)
\left[\textrm{erf}\left(\frac{1-\mu_0}{\beta}\right)
+\textrm{erf}\left(\frac{1+\mu_0}{\beta}\right)\right].
\end{equation}
Here, $\textrm{erf}(x)$ is the error function and $\textrm{D}_{-3}(z)$ is the parabolic cylinder function \cite{Gradshteyn2000}. Some contour plots of the electron distribution given by equation (1) with different parameters are shown in figure 1.

\subsection{General expression of the growth rate}

We assume that the thermal electrons are the dominant electron species in plasma, and a tenuous component of non-thermal electrons are the energy source to amplify the fast electromagnetic waves. According to the cold plasma theory, the dispersion relation for the extraordinary (X) and ordinary (O) modes is given by
%3
\begin{equation}
N_\sigma^2=\varepsilon_\sigma, \nonumber
\end{equation}
\begin{equation}
\varepsilon_X=1-\frac{\omega_{pe}^2}{\omega(\omega+\tau\Omega_e)}, \ \ \ \
\varepsilon_O=1-\frac{\tau\omega_{pe}^2}{\omega(\tau\omega-\Omega_e\cos^2\theta)},
\end{equation}
where $N_\sigma$ is the refractive index, the subscript $\sigma$ represents the wave mode, $\sigma=O$ is the ordinary (O) mode, and $\sigma=X$  is the extraordinary (X) mode, $\omega_{pe}$ and $\Omega_e$ are the ambient plasma frequency and electron-cyclotron frequency, $\omega$ and $\theta$  denote the wave angular frequency and the wave phase angle defined with respect to the ambient magnetic field, respectively. The parameter $\tau$ is defined as
\begin{equation}
\tau=\left(s+\sqrt{\cos^2\theta+s^2}\frac{\omega^2_{pe}-\omega^2}{\left|\omega^2_{pe}-\omega^2\right|}\right), \ \ \ \
s=\frac{\omega\Omega_e\sin^2\theta}{2\left|\omega^2_{pe}-\omega^2\right|}.
\end{equation}

The general expression of the temporal growth rate of ECM instability is given by \cite{Wu2002, Lee2013}
\begin{eqnarray}
\Gamma_\sigma & = & \frac{\pi}{2}\frac{n_b}{n_0}\frac{\omega_{pe}^2}{\omega}\frac{1}{(1+T_\sigma^2)R_\sigma}
\sum_{m=-\infty}^{\infty}\int{d^3\textbf{u}(1-\mu^2)\delta\left(\gamma-\frac{m\Omega_e}{\omega}-\frac{N_\sigma u\mu}{c}\cos\theta\right)}
\nonumber \\
\ & \ & \times \left\{\frac{\omega}{\Omega_e}\left[K_\sigma\sin\theta+T_\sigma\left(\cos\theta-\frac{N_\sigma u\mu}{c}\right)\right]
\frac{J_m(b_\sigma)}{b_\sigma}+J_m^\prime(b_\sigma)\right\}^2
\nonumber \\
\ & \ & \times \left[ u\frac{\partial}{\partial u} + \left(\frac{N_\sigma u}{c}\cos\theta-\mu\right)
\frac{\partial}{\partial\mu}\right] F_e(u,\mu),
\end{eqnarray}
with
\begin{eqnarray}
T_X & = & -\frac{\cos\theta}{\tau},\ \ \ \ \ \ \ \ \ \ \ \ \ \ \ \ \ \ \ \ \ T_O = \frac{\tau}{\cos\theta},
\nonumber \\
R_X & = & 1-\frac{\tau\omega_{pe}^2\Omega_e(1+U)}{2\omega(\omega+\tau\Omega_e)^2}, \ \ \ \ \
R_O = 1+\frac{\tau\omega_{pe}^2\Omega_e(1-U)\cos^2\theta}{2\omega(\tau\omega-\Omega_e\cos^2\theta)^2},
\nonumber \\
K_X & = & \frac{\omega_{pe}^2}{\omega^2-\omega_{pe}^2} \frac{\Omega_e\sin\theta}{\omega+\tau\Omega_e}, \ \ \ \ \ \ \
K_O = \frac{\omega_{pe}^2}{\omega^2-\omega_{pe}^2} \frac{\tau\Omega_e\sin\theta}{\tau\omega-\Omega_e\cos^2\theta},
\nonumber \\
U & = & \frac{\tau^2-\cos^2\theta}{\tau^2+\cos^2\theta} \frac{\omega^2+\omega_{pe}^2}{\omega^2-\omega^2_{pe}}, \ \ \ \
b_\sigma = \frac{\omega}{\Omega_e} N_\sigma u(1-\mu^2)^{1/2}\sin\theta,
\end{eqnarray}
where $\gamma=(1+u^2/c^2)^{1/2}$  is the relativistic factor and we have replaced all $\gamma$ outside the delta function by unity in the weakly relativistic approximation, $n_b$  and $n_0$ denote the number densities of non-thermal electrons and background electrons, $J_m(b_\sigma)$ and $J_m^\prime(b_\sigma)$ are the Bessel function of order $m$ and its first derivative, respectively. Here, we would like to point out that the electron-cyclotron maser works, in principle, only for relativistic distribution. The distribution function equation (1) is a simple Maxwellian distribution with pitch-angle anisotropy, which is reasonable for weak relativity. Otherwise, it is better to use the J\"{u}ttner distribution, which is the relativistically generalized classical Maxwell-Boltzmann distribution \cite{Juttner1911, Treumann2016}. This is beyond the scope of this paper.

\section{General properties of electron-cyclotron-maser instability}

One can calculate the growth rate of emission in X mode or O mode for a given electron distribution and the plasma-to-electron cyclotron frequency ratio $\omega_{pe}/\Omega_e$ on the basis of equations (1) and (5), which is a function of the wave frequency $\omega$  and the propagation phase angle $\theta$. In the present discussion, we investigate the fundamental and second harmonic O mode and X mode waves which are designated by O1, X1, O2 and X2 for abbreviation. In principle, the third or higher harmonics can also be excited with the increase of $\omega_{pe}/\Omega_e$, but their growth rates are much lower than the values of the fundamental and second harmonic wave modes\cite{Lee2013}.

Figure 2 shows the normalized growth rate $\Gamma/\Omega_e$ of wave modes O1, X1, O2 and X2, plotted as a function of the normalized wave frequency $\omega/\Omega_e$  and the phase angle $\theta$, for several values of $\omega_{pe}/\Omega_e$. In obtaining these numerical results, the input parameters in the distribution function are taken to be $u_0=0.3c$, $\alpha=0.1u_0$, $\beta=0.5$ and $\mu_0=0.5$ ($\phi_0=60^\circ$), namely, the distribution function shown in the right-top panel in figure 1 is used. Note that the frequency ratios $\omega_{pe}/\Omega_e$ are not same for the cases results shown in different panels in figure 2. For each wave mode from left to right, the first, second and third rows display the results for the case that $\omega_{pe}$ is much less than $\Omega_e$, the case of $\omega_{pe}/\Omega_e$ near which the wave is most unstable, and the case of $\omega_{pe}/\Omega_e$ that is near its upper limit for instability, respectively. The upper limit of $\omega_{pe}/\Omega_e$ for the excitation of each wave mode is determined by the necessary that its cutoff frequency is not much greater than the fundamental or harmonic electron-cyclotron frequency, which are approximately equal to 0.3, 1.0, 1.4 and 2.0 for modes X1, O1, X2 and O2, respectively.

One can see from figure 2 that both forward propagating waves ($\theta<90^\circ$) and backward propagating waves ($\theta>90^\circ$) can be excited for all wave modes, and the growth rates of the former are larger than that of the later. When the ratios $\omega_{pe}/\Omega_e$ are far away from their upper limit values for wave excitation, the most unstable waves propagate nearly perpendicular to the ambient magnetic field with phase angle $\theta\approx90^\circ$ for O1, O2 and X2 mode, while the most unstable wave of X1 mode propagates more obliquely with $\theta\lesssim80^\circ$. On the other hand, as $\omega_{pe}/\Omega_e$ approaches its upper limit for instability, every mode becomes to be quasi-parallel propagation.

Figure 3 plots the maximum growth rate $\Gamma_{max}/\Omega_e$  (upper panel), the frequency $\omega_{max}/\Omega_e$ (middle panel) at which the maximum growth occurs, and the wave propagation angle $\theta_{max}$ (bottom panel) corresponding to the maximum growths for each mode as a function of the frequency ratio $\omega_{pe}/\Omega_e$. The blue, red, black and green lines represent the results of X1, O1, X2 and O2 wave modes, respectively. As has been noted by many researchers, the maximum growth rates depend on the frequency ratio $\omega_{pe}/\Omega_e$ for each mode which can be unstable in a finite range of $\omega_{pe}/\Omega_e$. With the increase of $\omega_{pe}/\Omega_e$, modes X1, O1, X2 and O2 are stabilized in subsequence, and the propagation angle for all modes changes from quasi-perpendicular direction to quasi-parallel direction before the modes are suppressed. At the same time, the frequencies of each mode start very close to the harmonic cyclotron frequency $m\Omega_e$ ($m=1,2$) but systematically deviate away from the harmonic frequency as $\omega_{pe}/\Omega_e$ approaching its upper limit for instability.

These cases studies give us a brief survey on several basic properties of the ECM instability, which are generally similar for different models of electron velocity distribution shown in figure 1.

\section{Dependence of growth rate on the intensity of Alfv\'en waves}

To investigate the influences of Alfv\'en waves on the maser instability, we have done a comprehensive parametric calculation on the variations of the maximum growth rate with the parameters $\alpha$, $\beta$, $u_0$ and $\mu_0$ in the distribution function equation (1) for a given $\omega_{pe}/\Omega_e$.

As one can see from figure 3, the maximum growth rates of each wave mode first increase with increasing $\omega_{pe}/\Omega_e$ to a peak value, then sharply drop as $\omega_{pe}/\Omega_e$  approaching its upper limit for instability. The growth rates peak at $\omega_{pe}/\Omega_e\approx0.15$ , 0.7 for X1 and O1 modes, and at $\omega_{pe}/\Omega_e\approx1.1$  for both X2 and O2 modes. The variation tendencies of the maximum growth rate with   $\omega_{pe}/\Omega_e$ are generally similar for different electron velocity distributions. Thus, without loss of generality, we let $\omega_{pe}/\Omega_e$ equal to 0.15, 0.7, 1.1 and 1.1 for X1, O1, X2 and O2 modes respectively in what follows.

Firstly, we discuss the influence of Alfv\'en waves on the growth rate of maser instability excited by ring-beam electrons with different average pitch-angles. Contour plots of the maximum growth rate as function of $\phi_0$ and $\beta$ are presented in figure 4, where $\phi_0=\arccos\mu_0$  is the average pitch-angle, and $\beta^2\approx 2B_w^2/B_0^2$ is proportional to the magnetic field intensity of Alfv\'en waves. In these calculations, we choose $u_0=0.3 c$  and $\alpha=0.1 u_0$. The left-top, right-top, left-bottom and ring-bottom panels are results for O1, O2, X1 and X2 modes, respectively. The results indicate that the dependencies of the growth rate on the strength of Alfv\'en waves are not only different for different wave mode, but also are different for different average pitch-angle of the ring-beam electrons.

For the O1 mode, the growth rates increase with the increase of $\beta$ for pitch-angle $\phi_0\lesssim30^\circ$ and $\phi_0\gtrsim70^\circ$,  while the growth rates decrease with the increase of $\beta$ for $30^\circ<\phi_0<70^\circ$. These imply that whether the non-thermal electron distribution are predominant by the beam component or the electrons are nearly a pure ring, the pre-existing Alfv\'en waves will enhance the growth of the maser instability of O1 mode. If the amplitude of the ring and beam components are comparable with each other, the pre-existing Alfv\'en wave will weaken the growth of the maser instability. For a giving intensity of Alfv\'en waves, if the intensity is not very strong, for example $\beta<0.5$, the growth rates of ring-beam electrons with comparable two components are generally larger than that of beam-dominated electrons or pure ring electrons with the same electron energy. However, the growth rate of beam-dominated electrons becomes the largest one if the Alfv\'en wave intensity is strong enough with $\beta\gtrsim0.6$.

For the O2 mode, the growth rates increase with the increase of $\beta$ for $\phi_0<25^\circ$. This indicates that Alfv\'en waves enhance the maser excitation of O2 mode by beam-dominated non-thermal electrons, which is similar to the O1 mode. Meanwhile, the Alfv\'en waves have a weakening effect on the maser instability for $\phi_0>30^\circ$. Another interesting point is that the effects of Alfv\'en wave intensity on the O2 mode approach quickly to a saturated level at $\beta\approx0.3$.

For the X1 mode, the growth rates decrease almost always with the increase of $\beta$, except in a very small region near $\phi_0=90^\circ$ where the growth rates first increase slightly, then decrease with the increase of $\beta$. This means that the pre-existing Alfv\'en waves have a weakening effect on the growth of X1 mode in general.

For the X2 mode, the growth rates increase greatly with the increase of $\beta$ for pitch-angle $\phi_0\lesssim50^\circ$, while the growth rates decrease slowly with the increase of $\beta$ for other pitch-angles. These indicate that Alfv\'en waves enhance the maser excitation of X2 mode driven by beam-dominated electrons, but weaken the maser instability produced by ring-dominated electrons.

In brief summary, if the velocity distribution of non-thermal electrons for maser instability are dominated by a beam component, the pre-existing Alfv\'en waves will enhance the growth rate of instability for O1, O2 and X2 modes but weaken the growth rate of X1 mode. For non-thermal electrons with a ring distribution, Alfv\'en waves can also enhance the growth rate of the fundamental mode, but weaken the growth rate of the second harmonic mode. If the non-thermal electrons have a velocity distribution with comparable beam and ring components, Alfv\'en waves generally weaken the growth rate of all wave modes.

Next, figure 5 shows the contour plots of the maximum growth rate as function of the average pitch-angle $\phi_0$ and the average electron momentum  $u_0$. The other parameters used for these calculations are $\alpha=0.03c$ and $\beta=0.5$. The left-top, right-top, left-bottom and right-bottom panels are results for O1, O2, X1 and X2 modes, respectively. In general, the growth rates of all wave mode have an increasing trend with respect to the increase of electron momentum. For a fixed value of the average electron momentum (or energy), figure 5 seems to indicate that the growth rates of ring-dominated electrons are larger than that of beam-dominated electrons for X2 mode, while it is reversed for other wave modes. However, these are correct only for $\beta\gtrsim0.5$ when the spread of the electron pitch-angle is scattered wide enough by intense Alfv\'en waves. Otherwise, as one can see from figure 4, for the pitch-angle spread width $\beta\lesssim0.4$, the maximum growth rates peak at the region where the pitch-angle $\phi_0$ is about $40^\circ$to $60^\circ$ for O1, O2 and X1 modes.

Finally, we have a discussion on the effect of electron momentum dispersion $\alpha$ which can also be considered as the ``temperature'' of the ring-beam electrons. The lower the value of $\alpha$ is, the ``colder'' the ring-beam is. On the basis of equation (1), one can expect intuitively that the colder the ring-beam is, the higher the growth rate would be, since the gradient of distribution function in the momentum space increases with the decrease of $\alpha$. Indeed, this is consistent with the numerical results, which we don't show. An interesting result found from the numerical calculation is that the wave growth rates are more sensitive to the variation of Alfv\'en wave intensity for ``hotter'' ring-beam electrons than that for ``colder'' ring-beam electrons. For example, when the parameter value $\beta$ decreases from 1.0 to 0.1 with fixed parameter values $u_0=0.3 c$,  $\phi_0=60^\circ$ and $\omega_{pe}/\Omega_e=1.1$, the maximum growth rates of O2 mode increase 3.06 times and 23.2 times for $\alpha=0.01c$ and $\alpha=0.09c$ (corresponding to a ``temperature'' about $3.0\times10^5$ k and $2.4\times10^7$ k), respectively.

\section{CONCLUSION}

Large amplitude Alfv\'en waves intrinsically pervade the solar atmosphere and interplanetary space. Their roles on the heating and acceleration of solar corona and solar wind have been the hot and frontier research field for more than half centuries. Recently, it is suggested that Alfv\'en waves play important roles on the ECM instability \cite{Wu2006,Wu2012,Wu2014}. The electron-cyclotron maser is a process that can generate coherent radio radiation directly from non-thermal electrons in magnetized plasma, which has been one of the dominant mechanism for producing high-power radio emissions observed in our universe \cite{Treumann2006}. In this paper, we present a comprehensive parametric investigation on the influences of Alfv\'en waves on the cyclotron-maser instability driven by non-thermal ring-beam electrons.

It is found that both forward propagating and backward propagating waves can be excited in the X mode and O mode near the fundamental and harmonic electron-cyclotron frequencies in a finite range of frequency ratio $\omega_{pe}/\Omega_e$. The X1 mode propagates more obliquely with respect to the ambient magnetic field while O1, O2 and X2 modes propagate quasi-perpendicularly in general. However, each mode becomes to be quasi-parallel propagation as the frequency ratio $\omega_{pe}/\Omega_e$  approaches its upper limit for instability.

Whether the pre-existing Alfv\'en waves enhancing or weakening the instability depend not only on the wave mode but also on the relative amplitude of the ring and beam components in the electron velocity distribution. The growth rates of X1 mode are weakened by Alfv\'en waves in general. For a beam-dominated distribution (the electron average pitch-angle $\phi_0\lesssim30^\circ$), the growth rates of maser instability for O1, O2 and X2 mode are enhanced with the increase of Alfv\'en wave energy density. In the other conditions, the growth rate of O1, O2 and X2 mode weaken with the increasing Alfv\'en wave intensity, except that the growth rate of O1 mode may also be enhanced by Alfv\'en waves for a ring distribution. In addition, the growth rates are more sensitive to the variation of Alfv\'en wave intensity for ``hotter'' ring-beam electrons than that for ``colder'' ring-beam electrons.

An implicit assumption in this study is that the electron distribution function does not change with time. In real situation, the electron velocity distribution may evolve with time. For instance, when the non-thermal electrons travel along the magnetic field line with decreasing background magnetic field strength, their average pitch-angle will decrease due to the conservation of magnetic moment, so that the electrons may evolve from a ring-predominant distribution to a beam-dominated distribution, or vice versa; the non-thermal electrons may lose some of their energy via collisions with the ambient plasma; and the energy density of Alfv\'en waves experienced by the evolving electrons may also vary with position and time. All of these can affect the spectral intensity, polarization, directivity and other properties of the radio emission amplified by non-thermal electrons through the cyclotron-maser instability, so the above results is important for us to understand the mechanism for the radio bursts with various fine structures observed in space and astrophysical plasmas.

\section*{ACKNOWLEDGMENTS}

The research is supported by the National Nature Science Foundation of China (41574167, 41421063) and the Fundamental Research Funds for the Central Universities (WK2080000077).

%\newpage

\clearpage
%%%%%%%%%%%%%%%%%
\begin{figure}[h]
\includegraphics[width=37pc]{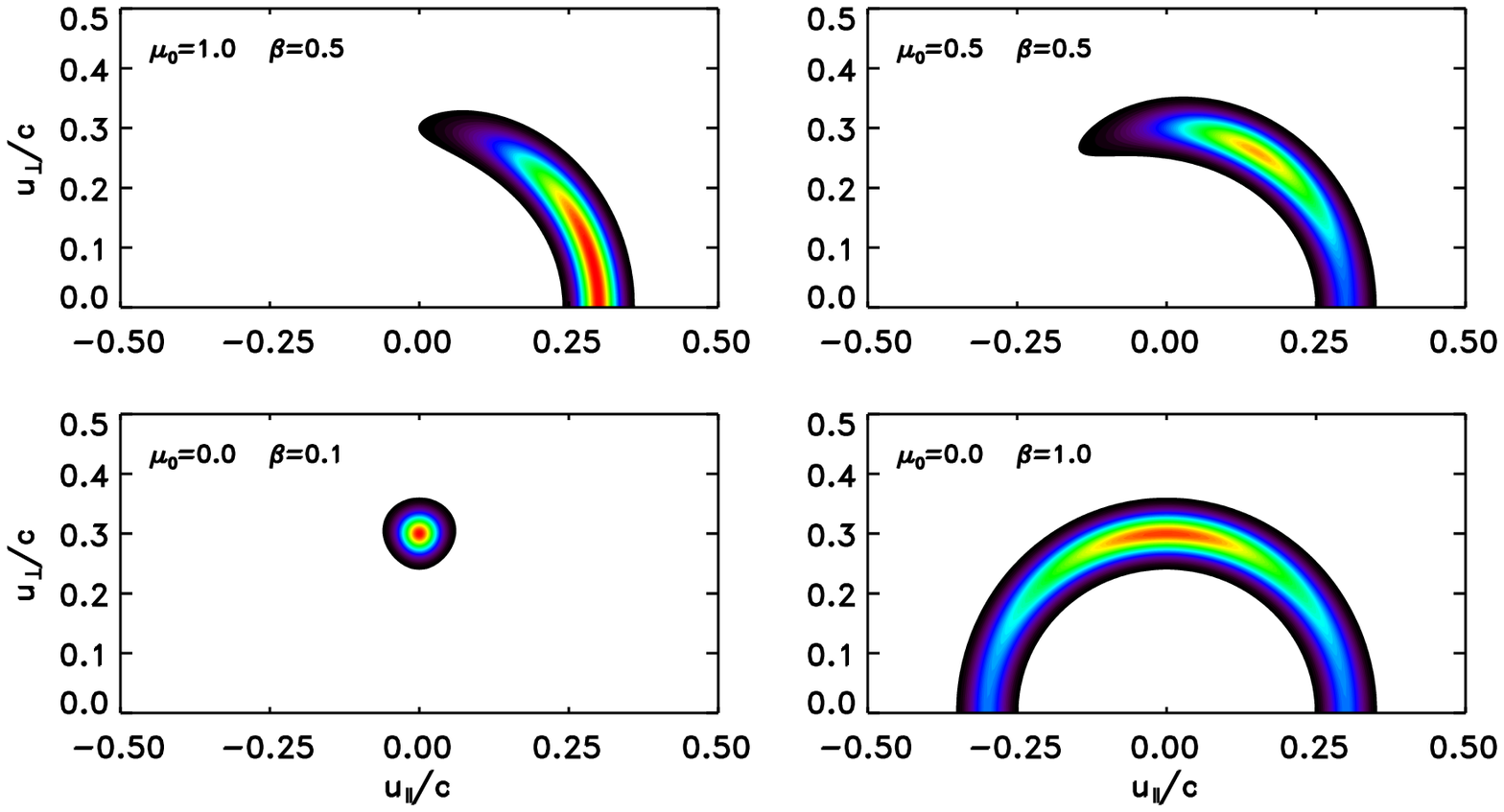}
\caption{Contour plots of the ring-beam non-thermal electron distribution function $F_e(u,\mu)$  versus $u_\perp$  and $u_\parallel$
in momentum space for different parameter values of $\mu_0$ and  $\beta$, where $u_0=0.3 c$ and $\alpha=0.1 u_0$ are used for the plot in all panels. }
\label{F1}
\end{figure}

\clearpage
%%%%%%%%%%%%%%%%%
\begin{figure}[h]
\includegraphics[width=34pc]{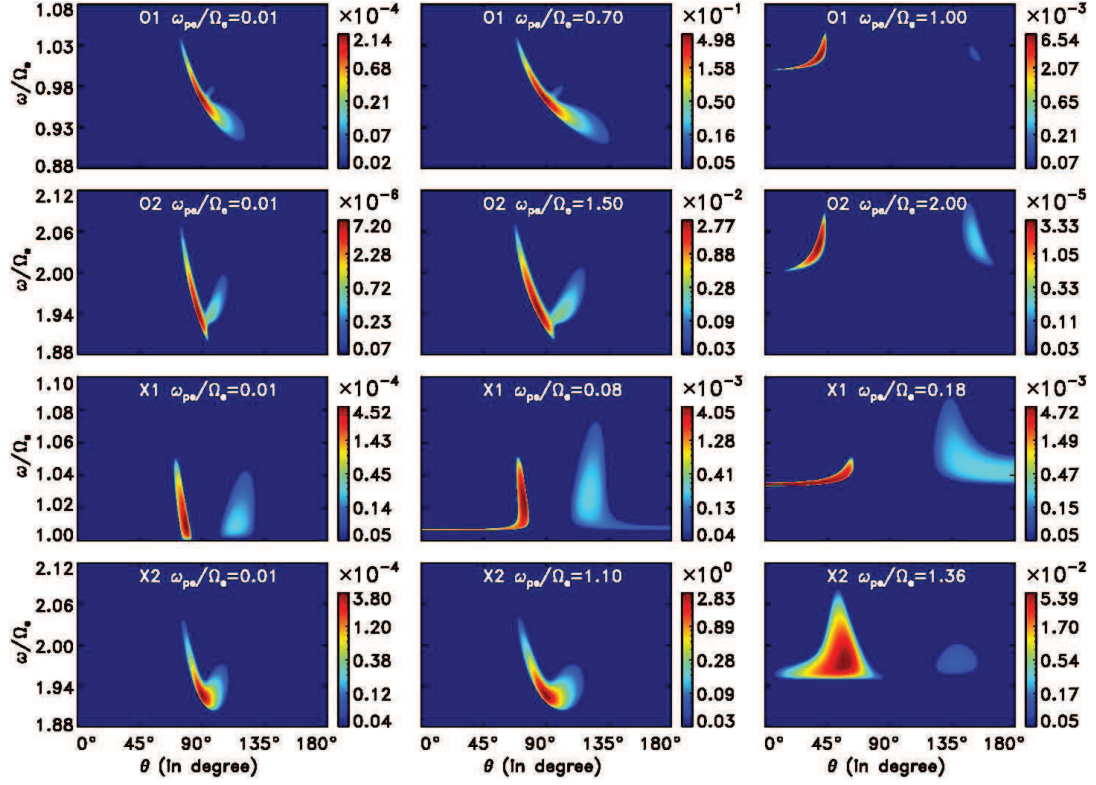}

\caption{Contour plots of the growth rate versus the wave propagation angle $\theta$ and wave frequency $\omega/\Omega_e$ for the fundamental and harmonic X modes and O modes at different frequency ratio $\omega_{pe}/\Omega_e$ for $u_0=0.3c$, $\alpha=0.1u_0$, $\beta=0.5$ and $\mu_0=0.5$.}
\label{F2}
\end{figure}

\clearpage
%%%%%%%%%%%%%%%%%
\begin{figure}[h]
\includegraphics[width=20pc]{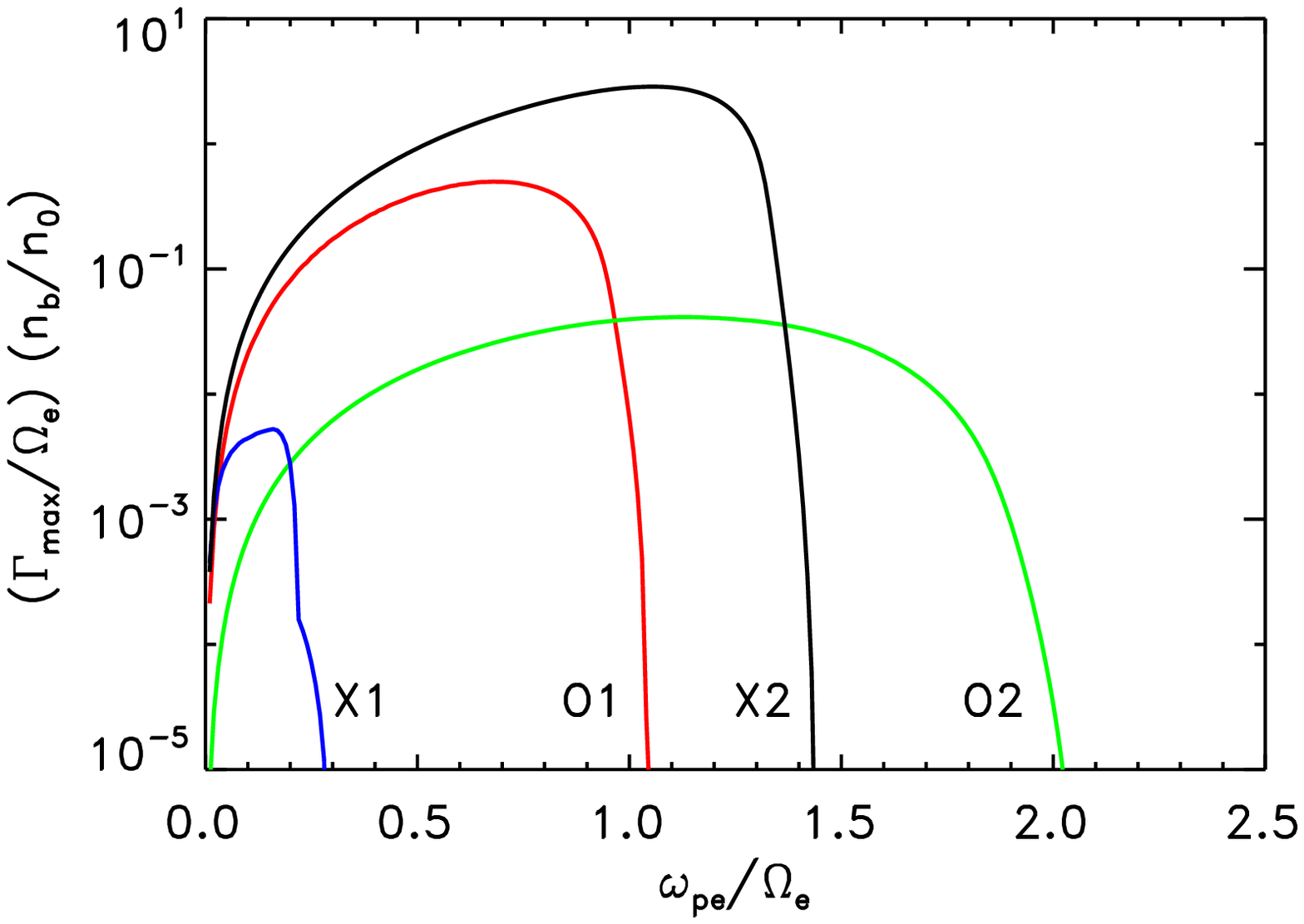}
\includegraphics[width=20pc]{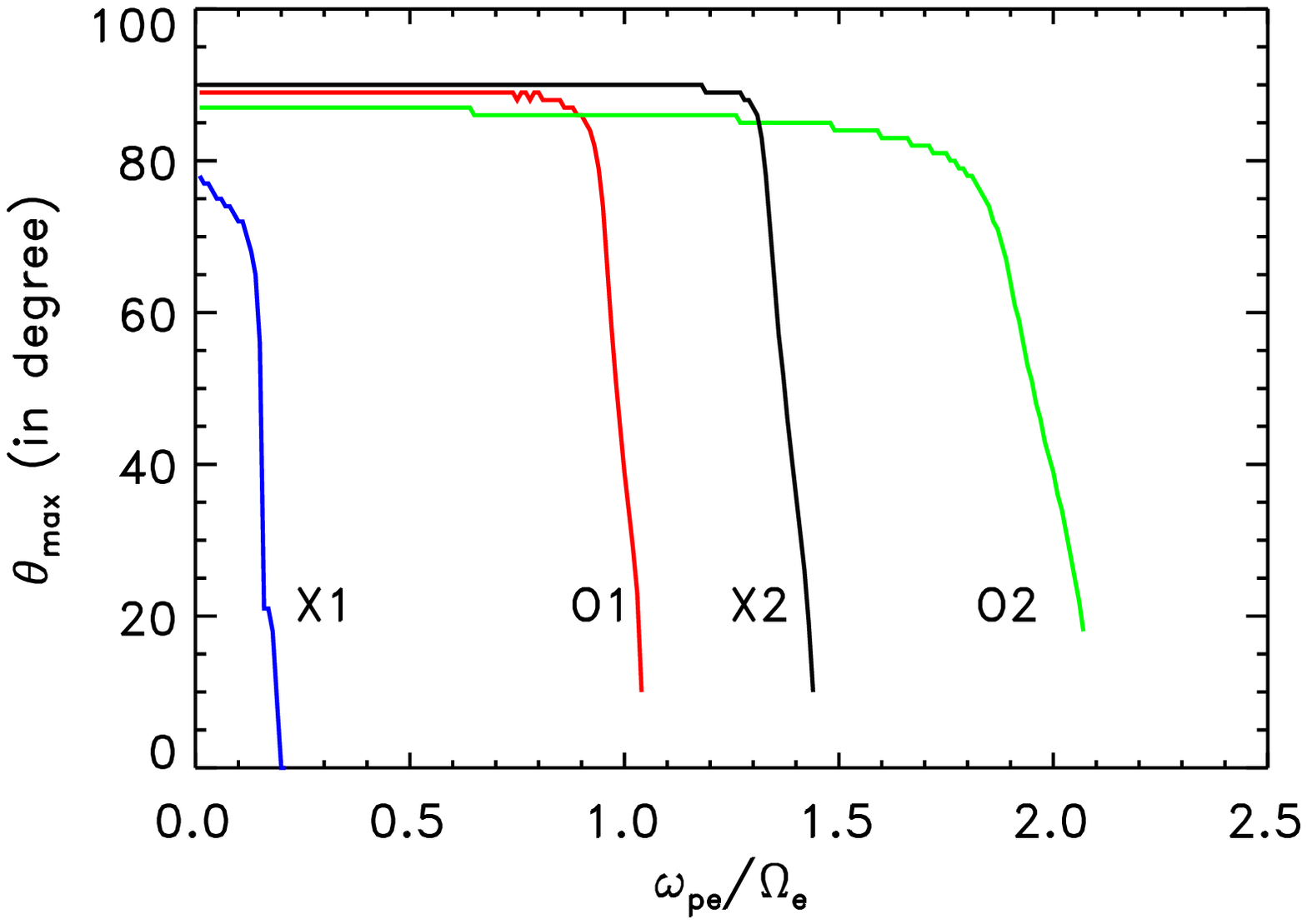}
\includegraphics[width=20pc]{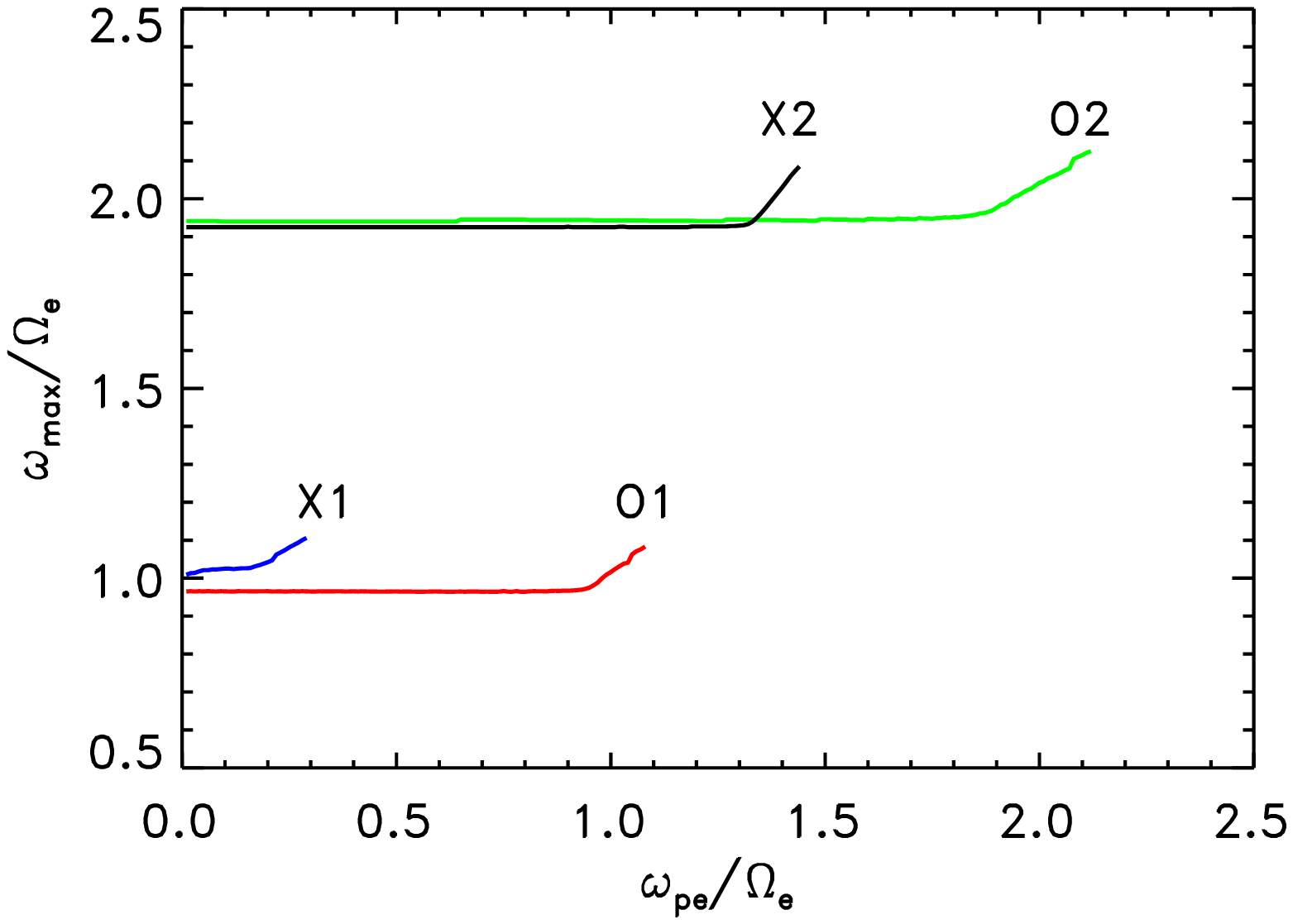}
\caption{The plot of the maximum growth rate $\Gamma_{max}/\Omega_e$, propagation angle $\theta_{max}$, and wave frequency $\omega_{max}/\Omega_e$ as a function of the frequency ratio $\omega_{pe}/\Omega_e$ for $u_0=0.3c$, $\alpha=0.1u_0$, $\beta=0.5$ and $\mu_0=0.5$.}
\label{F3}
\end{figure}

\clearpage
%%%%%%%%%%%%%%%%%
\begin{figure}[h]
\includegraphics[width=37pc]{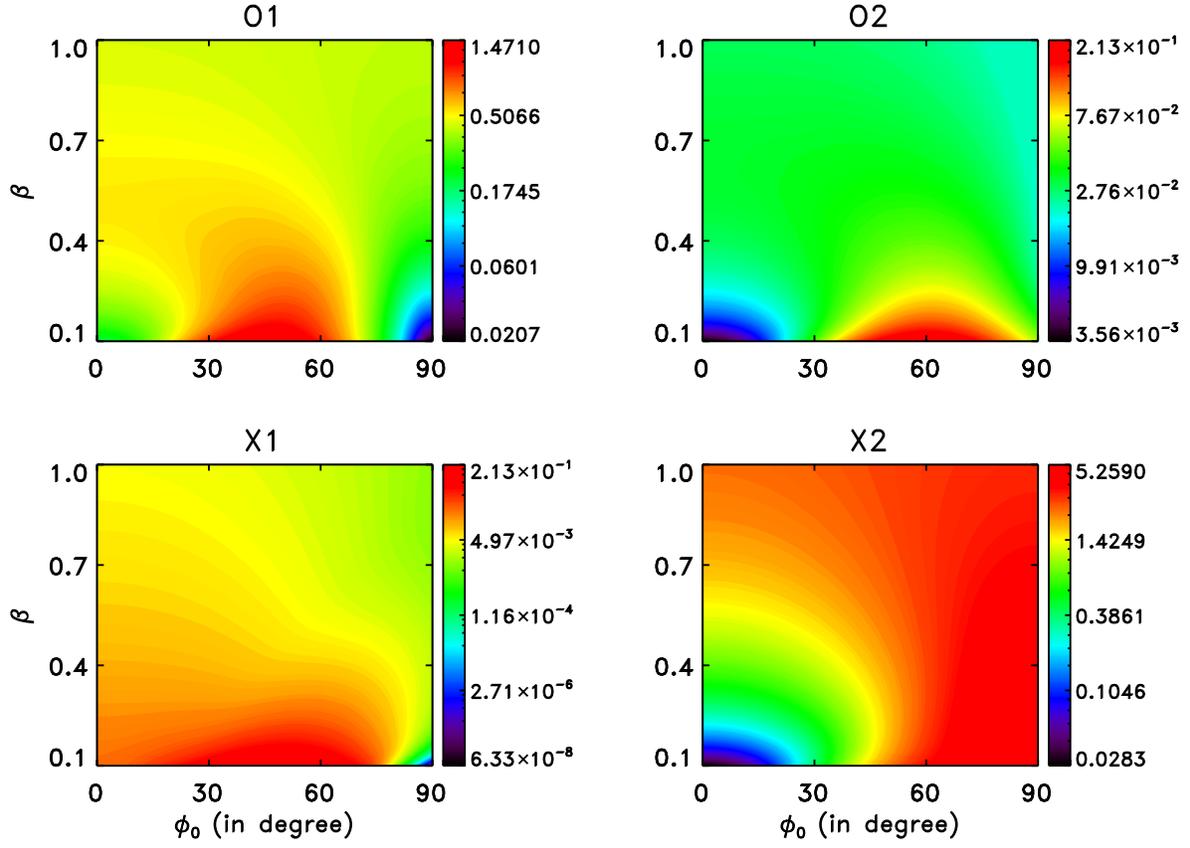}
\caption{Contour plots of the logarithm of the maximum growth rate versus electron average pitch-angle $\phi_0$ and the spread width of pitch-angle $\beta$ for $u_0=0.3c$ and $\alpha=0.1u_0$. Note that $\beta$ is proportional to the magnetic field intensity of Alfv\'en waves.}
\label{F4}
\end{figure}

\clearpage
%%%%%%%%%%%%%%%%%
\begin{figure}[h]
\includegraphics[width=37pc]{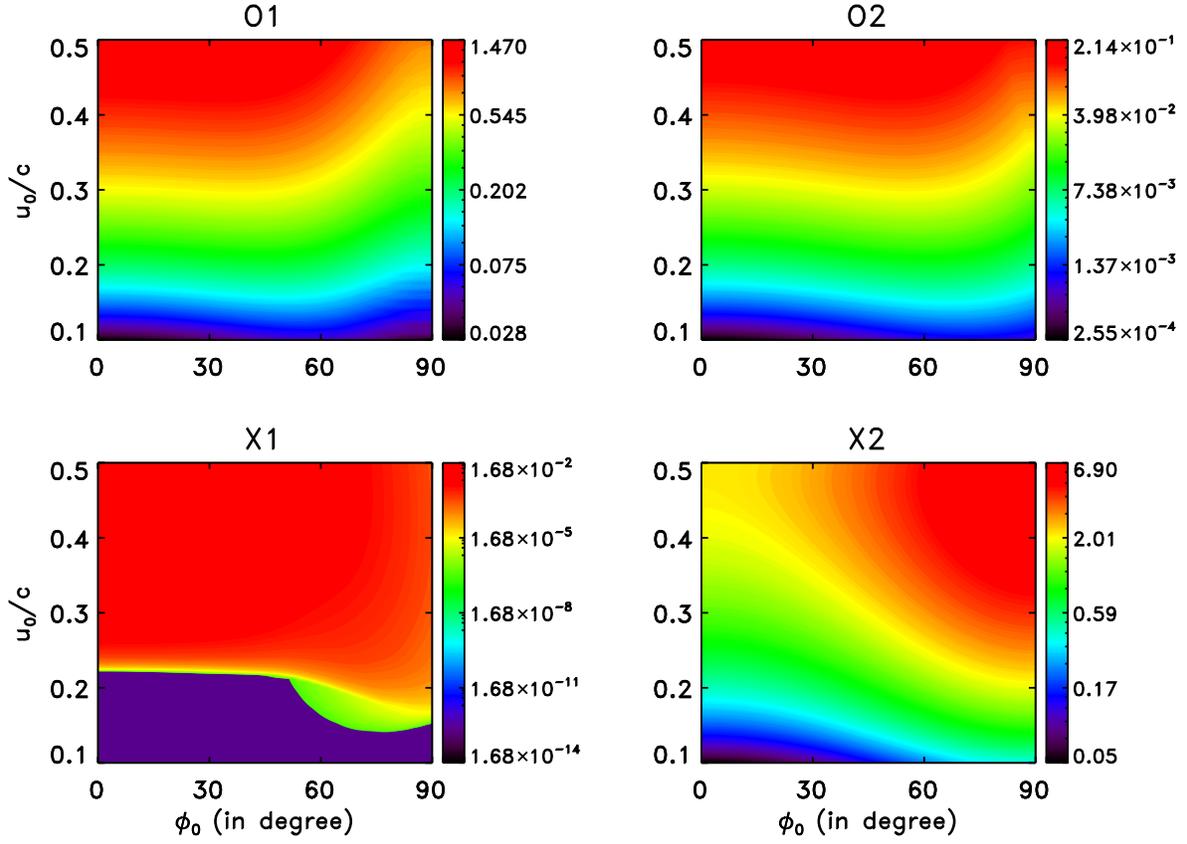}
\caption{Contour plots of the logarithm of the maximum growth rate versus average pitch-angle $\phi_0$ and average electron momentum $u_0$ for $\alpha=0.03c$ and $\beta=0.5$.}
\label{F5}
\end{figure}

\clearpage

\end{document}